# Phase transition in oxygen-intercalated pseudocapacitor Pr$_2$MgMnO$_6$ electrode: A combined structural and conductivity analysis

Moumin Rudra[1,*], S. Saha[2] and T. P. Sinha[1]

[1]*Department of Physics, Bose Institute, 93/1, A. P. C. Road, Kolkata, India 700009.*

[2]*Department of Physics, Oakland University, Rochester, Michigan, USA 48309.*

*Email: *iammoumin@gmail.com*

**Abstract**

The phase transition behavior and charge storage mechanism of Pr$_2$MgMnO$_6$ (PMM), an oxygen-intercalated pseudocapacitor, were investigated through crystal structure analysis, Raman spectroscopy, ac conductivity spectroscopy, X-ray photoelectron spectroscopy, and electrochemical spectroscopy. The crystal structure analysis and vibration studies revealed a phase transition in PMM, following the sequence *P2$_1$/n* → *I4/m* → *Fm$\bar{3}$m*. High-temperature Raman spectroscopy demonstrated a significant feature of a monoclinic-to-tetragonal phase transformation in PMM. The ac conductivity spectroscopy exhibited a semiconductor-to-metal transition in PMM. X-ray photoelectron spectroscopy of the Mn *2p* state confirmed the presence of oxygen vacancies in PMM at room temperature. Furthermore, the electrochemical performance of PMM as an electrode was evaluated. The PMM electrode displayed an intercalated pseudocapacitive nature, exhibiting a maximum possible specific capacitance of 257.57 F/g. The charge storage process of the PMM electrode was thoroughly reviewed and discussed, shedding light on the underlying mechanisms.

## 1. Introduction

A supercapacitor serves as a large-capacitive capacitor that bridges the gap between conventional capacitors and rechargeable batteries [1-8]. However, its limited power capability hampers its widespread applicability. To address this drawback, numerous projects have focused on designing electronic devices for high-speed supercapacitors [9-11]. These electronic devices can be categorized into two groups based on their charge storage mechanisms: double-layer capacitors [DLC] [12] and pseudo-capacitors [13]. DLC materials store charges without electrochemical reactions, with charges accumulating at the electrode-electrolyte interface. Conversely, pseudo-capacitor materials store charge through redox reactions or ion intercalation between the electrolyte and the electrode surface. The latter, known as the pseudocapacitor, exhibits both high power and high energy density. Various electrode materials, including polymers, carbon-based materials, hydroxides, and transition metal oxides, have been investigated [14-26], with transition metal oxides demonstrating superior electrochemical stability and durability.

Transition metal-based perovskite oxides have emerged as promising candidates for supercapacitor applications due to their thermal and chemical stability as well as their ability to accommodate oxygen vacancies. In 2014, Mefford et al. [27] reported on an oxygen-intercalated pseudocapacitor charge storage mechanism in LaMnO3. More recently, Liu et al. [28] discovered excellent oxygen-based intercalated pseudocapacitive behavior in the Pr-based double perovskite PrBaMn2O6. Inspired by these findings, we aim to explore charge storage mechanisms in $Pr_2MgMnO_6$ (PMM).

The investigation of structural transitions in double perovskite oxides is of paramount importance, particularly the tetragonal-monoclinic phase transition [29, 30]. Most oxide materials with perovskite structures are reported in the monoclinic *$P2_1/n$* space group [31] (occasionally in *I2/m* [32] and *$P2_1/m$* [33]) or the cubic *$Fm\bar{3}m$* space group [31]. Fewer materials with tetragonal symmetry have been identified, primarily in *I4/m* [34], but also in *I4/mm* [35, 36], *I4mm* [37], *$P4_2/n$* [38], *$P4_2/m$* [39], rhombohedral *$R\bar{3}m$* [40], *$R\bar{3}$* [32], orthorhombic *Pmm2* [41, 42], and even triclinic *$I\bar{1}$* [43]. Publications regarding the phase transitions of these phases are limited [44]. The *$P2_1/n$* → *I4/m* → *$Fm\bar{3}m$* phase transition sequence is exceedingly rare in double perovskites [29, 30]. In this article, we present X-ray diffraction results elucidating the crystal structure and phase transition process in PMM.

The successful synthesis of PMM is reported for the first time in this article. X-ray photoelectron spectroscopy was employed to investigate the presence of oxygen vacancies in PMM. The electrical properties of PMM, including electrical conductivity, and the electrochemical charging-discharging behavior were investigated using electrochemical impedance spectroscopy. The potential applications of PMM in both DLC and pseudo-capacitor systems are also explored. This new material, PMM, opens up exciting avenues for electrochemical applications in the years to come.

## 2. Experimental

PMM powder was synthesized using Pr$_2$O$_3$ (Sigma-Aldrich, 99.9%), MgO (Merck, 99%), and MnO$_2$ (LobaChemie, 99.99%) as starting materials. The conventional solid-state reaction method was employed, where the powders were calcined at 1573 K in air for 14 hours. The resulting calcined powders were then mixed with 5% polyvinyl alcohol, pressed into cylindrical pellets using a pelletizer, and sintered at 1623 K to obtain dense PMM pellets with a thickness of 1 mm and a diameter of 8 mm.

The crystal structure of PMM was investigated using a Rigaku Miniflex II X-ray powder diffractometer. The X-ray diffraction (XRD) measurements were conducted step by step at room temperature within the 2θ range of 10° to 120°. The collected data were processed using the Rietveld method with the FULLPROF software [45, 46]. During the XRD pattern refinement, a polynomial function with 6 coefficients was utilized to account for the background, and the peak morphologies were described by pseudo-Voigt functions. The microstructure of the PMM sample was examined using a scanning electron microscope (SEM) (FEI Quanta 200). For the Raman spectroscopy analysis, the room temperature Raman spectrum was acquired using a LABRAM HR 800 system equipped with an 1800 grooves/mm diffraction grating and a Peltier-cooled charge-coupled device (CCD) detector. An Ar-ion laser with a wavelength of 488 nm was employed to excite the sample, and the laser light was focused on the sample using a 100X objective lens with a numerical aperture (NA) of 0.9. X-ray photoelectron spectroscopy (XPS) measurements were conducted using an MXPS system (Scienta Omicron, Germany) equipped with an EA 125 U5 hemispherical analyzer, an XM1000 monochromator, and a CN10 charge neutralizer. Monochromatized Al-K radiation (1486.7 eV) with an output power of 300W was used for the XPS investigation. The binding energies (BE) of various species were determined using the C(*ls*) signal (284.9 eV) as a reference. XPSPEAK4.1 software was employed for data analysis, and XPS peak areas were used to calculate the atomic concentrations in the samples.

Impedance spectroscopy was performed on sintered PZM particles using an LCR meter (HIOKI-3532, Japan) in a temperature range of 300 K to 700 K. The oven temperature was controlled with an accuracy of 1K using a Eurotherm 818p programmable temperature controller. The measurements were conducted with an oscillating voltage of 1.0 V over a frequency range of 42 Hz to 5 MHz. Prior to testing, the flat surfaces of the samples were carefully cleaned, and a thin layer of silver was applied for electrical contact. The evaporation effect of the silver paste was monitored before and after the tests. The ac electrical conductivity (σ) was calculated from the conductance (G), where σ = Gd/A, and C$_0$ = ε$_0$A/d represented the empty cell capacitance, with A as the sample area and d as the sample thickness. The dc conductivity values of the PZM electrodes were obtained using the I-V characteristic measurements with a Keithley Data Acquisition Multimeter System. For the room temperature electrochemical characterization of the PMM electrodes, several techniques were employed. Cyclic voltammetry (CV) was performed to investigate the electrochemical behavior of the PMM electrodes. The measurements involved scanning the potential within a range of -0.2 to 0.6 V at various scan rates (5, 10, 20, 50, and 100 mV/s). This technique provides information about the redox reactions and charge storage capacity of the electrode material. To prepare the PMM electrodes, a gel was formed by mixing graphite powder (10 wt%), polyvinylidene fluoride (PVDF, 5 wt%), and PMM powders (85 wt%) in N-Methyl-2-pyrrolidone (NMP) under sonication. The gel was then drop-cast onto one flat surface of a Nickel foam substrate and dried in a vacuum at 343 K for 8 hours. The resulting PMM loading density on the Nickel foam was approximately 1 mg/cm². These experimental techniques and analyses provide valuable insights into the synthesis, crystal structure, microstructure, and electrical properties of PMM, contributing to a comprehensive understanding of its potential for electrochemical applications.

3. Results and Discussions
   1. Structural Analysis
   a. X-ray diffraction

In Fig. 1, the XRD profile of PMM at room temperature (RT) is displayed. The observed data is represented by black circles, while the calculated data is shown as a green line. The blue line represents the differences between the observed and calculated data, which appears as a straight line, indicating a good refinement. The reliability factors $R_p = 2.91$, $R_{wp} = 3.89$, $R_{exp} = 2.52$, and $\chi^2 = 2.39$ provide further confirmation of the quality of the refinement. The refined parameters at RT are listed in Table 1. The presence of the superlattice peak (101) at $20^0$ confirms the monoclinic $P2_1/n$ structure of PMM. The cell parameters obtained from the refinement are $a = 5.44826$ (5) Å, $b = 5.46948(4)$ Å, $c = 7.70182(5)$ Å, $\beta = 89.9063°$. A schematic diagram of the PMM unit cell is shown in the appendix of Figure 1. The unit cell consists of four types of atomic positions: *4e* (xyz) for $Pr^{3+}$, *2c* (½00) for $Mg^{2+}$, *2d* (½0½) for $Mn^{4+}$ and *4e* (xyz) for $O^{2-}$. The $Mg^{2+}$ and $Mn^{4+}$ ions are coordinated by six $O^{2-}$ ions to form octahedral $MgO_6$ and $MnO_6$ units, respectively. During the refinement process, there was no mixing observed between the two B-site cations

(Zn and Mn), which was confirmed by replacing Zn with Mn and vice versa. Additionally, the tolerance factor (*t*) can be calculated using Equation 1.

$$t = (r_{Pr} + r_O)/\sqrt{2}(r_B + r_O) \tag{1}$$

where $r_{Pr}$, $r_O$, and $r_B$ are the ionic radii of praseodymium (Pr), oxygen (O), and B-site cations, respectively. A tolerance factor value of 0.838 was obtained, further supporting the low symmetrical monoclinic phase of PMM.

The high-temperature X-ray diffraction (XRD) patterns of PMM, collected between room temperature (RT) and 573 K, are presented in Figure 2. It is observed that certain separation peaks, specifically at 2θ values of 40.4° and 46.8°, decrease in intensity as the temperature increases. This decrease indicates a reduction in unit cell distortion, suggesting a possible phase transition in PMM. At room temperature (303 K), the crystal structure of PMM is defined in the monoclinic *P2₁/n* space group. However, at higher temperatures (573 K), the structure can be identified in a tetragonal *I4/m* unit cell. The refined parameters for the crystal structure at 573 K are provided in Table 1. The Rietveld refinement of the XRD pattern for PMM at RT and 573 K, corresponding to different 2θ values, is shown in Figure 3. In Figure 3(a), the peak intensity of (111) at 2θ = 25.8° decreases with increasing temperature. Figure 3(b) illustrates that three peaks, namely (022), (202), and ($\bar{2}$02), merge into a single peak (022) at 2θ = 40.4°. Similarly, Figure 3(c) demonstrates that two peaks, (220) and (004), at 2θ = 46.8° for RT, merges into single peaks at 2θ = 46.7° for the temperature of 573 K. and Figure 3(d) shows three peaks, (224), (400), and ($\bar{2}$24) for RT merges to single peak at 68.6°. These observations strongly indicate a transformation from the lower-symmetry *P2₁/n* phase to the higher-symmetry *I4/m* phase. The refined lattice parameters obtained from the XRD data at different temperatures are presented in Figure 4. The plot clearly demonstrates that as the temperature increases, there is an increase in symmetry. In Figure 4(a), it can be observed that the cell parameters, *a*, *b*, and *c*, become approximately equal at higher temperatures, indicating a cubic $Fm\bar{3}m$ phase with the highest level of symmetry. Furthermore, Figure 4(b) depicts the transition behavior of PMM from the *P2₁/n* phase to the *I4/m* phase. It is observed that around 500 K, the *P2₁/n* phase begins to convert into the *I4/m* phase. This indicates a continuous phase transition in PMM, where the crystal structure evolves from the initial monoclinic *P2₁/n* phase at RT to the tetragonal *I4/m* phase at elevated temperatures. These findings suggest that the phase transition in PMM is not limited to the *P2₁/n* to *I4/m* transition but may potentially extend further to a higher symmetric cubic $Fm\bar{3}m$ phase at even higher temperatures. The temperature-dependent evolution of the crystal structure in PMM demonstrates a continuous transformation, highlighting the dynamic nature of the material and its ability to adopt different crystal symmetries as the temperature changes. The thermal variation of PMM reveals phase transitions occurring in the sequence *P2$_1$/n → I4/m → Fm$\bar{3}$m*. This observed phase transition sequence is consistent with the commonly observed transitions in ordered perovskites [29, 30]. Similar sequences have also been reported in certain fluoride elpasolites [47].

The grain-size distribution curve of PMM, which has been calcined at 1573 K, is presented in Figure 5. The grain-size distribution curve of PMM shows the range and distribution of grain sizes within the material. It provides information about the size distribution of individual grains in the PMM sample. The curve typically plots the grain size on the x-axis and the counts of grains on the y-axis. In the case of PMM, the grain-size distribution curve reveals that the grain sizes range from 0.4 µm to 0.75 µm. The average grain size is determined to be approximately 0.54 µm. This information helps in understanding the microstructure of PMM and provides insights into its physical properties and potential applications. The Scanning Electron Microscopy (SEM) images included in the appendix of the study showcase the surface morphology of PMM and reveal the presence of grains with different sizes. SEM is a powerful technique that allows for detailed examination of the sample's microstructure and surface features. The SEM images provide visual evidence of the variation in grain sizes within PMM. Grains refer to individual crystalline regions or particles within the material. The images show that PMM consists of grains of different sizes, indicating a heterogeneous microstructure. The presence of grains with different sizes suggests variations in growth or nucleation processes during the synthesis or fabrication of PMM. These images contribute to a comprehensive understanding of PMM's morphology and can assist in further investigations and optimization of its properties for specific applications.

b. **Raman Spectroscopy**

In Figure 6(a), the room temperature (RT) non-polarized Raman spectrum of the PMM electrode is shown. The spectrum exhibits three main peaks located at 450, 550 and 850 cm$^{-1}$. To enhance the visibility of smaller peaks, background correction techniques were applied, resulting in the RT Raman spectrum of PMM shown in Figure 6(b). This corrected spectrum reveals several smaller peaks in the 75-400 cm$^{-1}$ range, indicating the lower symmetry of the PMM crystal structure. As discussed in Section 3.1.a, the crystal structure of PMM at RT is characterized by low symmetry monoclinic $P2_1/n$. Within the $P2_1/n$ unit cell, there are a total of 60 phonon modes, out of which 24 are Raman active. The active Raman modes can be expressed using Equation 2, supported by factor group analysis [48].

$$\Gamma(P2_1/n) = 6T + 6L + 6\nu_5 + 4\nu_2 + 2\nu_1 \qquad (2)$$

According to factor group analysis, the Raman modes in the $P2_1/n$ unit cell include six translational modes ($T$), six vibrational modes of Pr-cations ($L$), six bending modes of MnO6 octahedra ($\nu_5$), four antisymmetric stretching modes of $MnO_6$ octahedra ($\nu_2$), and two symmetric stretching modes of $MnO_6$ octahedra ($\nu_1$). In the RT Raman spectrum, the peaks at 85, 107, 125, 140, and 188 cm$^{-1}$ (labeled as peaks 1, 2, 3, 4, and 5 in Figure 6(c)) correspond to the $T$ modes of Pr-cations. The peaks at 250, 293, and 372 cm$^{-1}$ (labeled as peaks 6, 7, and 8) are assigned as the $L$ modes of Pr-cations. The peak at 450 cm$^{-1}$ (peak 9) represents the $\nu_5$ mode of $MnO_6$ octahedra. Additionally, the peaks at 570, 613, and 741 cm$^{-1}$ (labeled as peaks 10, 11, and 12 in Figure 6(d)) correspond to the $\nu_2$ modes, while the peak at 813 cm$^{-1}$ (peak 13) represents the $\nu_1$ mode of $MnO_6$ octahedra. Exactly, out of the 24 Raman modes present in the $P2_1/n$ crystal structure of PMM, 13 modes are observed in the room temperature (RT) Raman spectrum. This discrepancy can be attributed to the small correlation field splitting between the different crystalline orientations within the polycrystalline PMM structure. In polycrystalline materials, the presence of multiple crystal orientations leads to a distribution of strain and local variations in the crystal environment. As a result, the Raman modes associated with specific crystal orientations may exhibit splitting or broadening due to these local variations. This effect is known as correlation field splitting. In the case of PMM, the small correlation field splitting between the different crystalline orientations causes the observed Raman modes to be broadened or split into multiple peaks.

Figure 7(a) presents the thermal variation of the Raman spectra of PMM, covering a temperature range from 303 K (room temperature, RT) to 673 K. The behavior of the Raman peaks indicates a phase transition occurring in PMM as the temperature increases. The PMM structure at RT is monoclinic $P2_1/n$, which deviates from the cubic $Fm\bar{3}m$ structure. The phase transition in PMM follows the sequence $P2_1/n \rightarrow I4/m \rightarrow Fm\bar{3}m$. This means that the crystal structure undergoes changes from monoclinic to tetragonal and finally to cubic symmetry. The observed disappearance of the major Raman peaks as the temperature rises suggests a transformation in the $P2_1/n$ structure. At higher temperatures, specifically at 673

K, the Raman spectrum of PMM exhibits a linear feature accompanied by increased noise. This linear feature is associated with the transition from the tetragonal to the cubic symmetry. The presence of high noise can be attributed to the small correlation field splitting between the different phases of PMM at elevated temperatures. To provide further insights into the correlation between the Raman modes of the monoclinic, tetragonal, and cubic phases of PMM, Table 2 is included, which presents a tabulated correlation between the phases. This table aids in understanding the changes in the vibrational modes and Raman peaks that occur during the phase transition process. Overall, the thermal variation of the Raman spectra of PMM depicted in Figure 7(a) illustrates the phase transition behavior and the corresponding changes in the crystal symmetry. The disappearance of Raman peaks and the appearance of a linear feature at high temperatures provide valuable information about the structural transformations occurring in PMM with increasing thermal energy.

Figure 7(b) and Figure 7(c) present the temperature dependencies of the symmetric stretching and linewidth of the $v_1$ mode of MnO$_6$ octahedra, respectively. The Raman frequency of the $v_1$ mode decreases as the temperature increases, indicating the involvement of anharmonic effects. The anharmonic model can be used to fit the Raman frequency and linewidth of the $v_1$ mode, as described by Equation 3 and Equation 4, respectively. These equations are supported by previous studies [49-54].

$$v = v_0 - A[1 + \frac{2}{e^x - 1}] \quad (3)$$

$$\Gamma = B[1 + \frac{2}{e^x - 1}] \quad (4)$$

where $A$ and $B$ are constants and $x = hv_0/2k_BT$, $v_0$ is the Raman frequency at 0 K, $k_B$ be the Boltzmann constant and T is the absolute temperature. The $v_0$ value from the fitting data was found to be 819 cm$^{-1}$. Overall, the thermal variation of the Raman spectra provides insights into the phase transition behavior of PMM, highlighting the changes in crystal structure and the involvement of anharmonic effects in the Raman-active modes. Understanding the changes in crystal structure and the involvement of anharmonic effects in the Raman-active modes is crucial for characterizing the properties and behavior of PMM. It provides valuable information about the structural stability, phase transitions, and lattice dynamics of the material. This knowledge is essential for designing and optimizing the performance of PMM and related materials in various applications.

### c. X-ray Photoelectron Spectroscopy

To investigate the presence of oxygen in the sample, we obtained the XPS spectrum of the Mn *2p* state and depicted it in Figure 8. The figure displays the deconvoluted peaks of Mn *2p_{3/2}* and Mn *2p_{1/2}* fitted with Lorentzian distributions, indicating the presence of mixed valence states of Mn. Peaks at 641.2 eV [53] and 653.4 eV [54] correspond to the $Mn^{3+}$ state, while major peaks at 641.6 eV [55] and 653.9 eV [56] indicate the $Mn^{4+}$ state. The presence of a small amount of $Mn^{3+}$ state suggests the existence of oxygen to maintain charge neutrality in the sample, as described by Equation 5 [57, 58]:

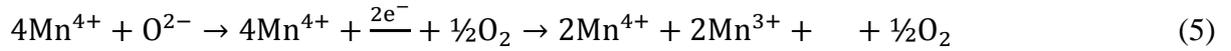

$$4Mn^{4+} + O^{2-} \rightarrow 4Mn^{4+} + \underline{2e^-} + ½O_2 \rightarrow 2Mn^{4+} + 2Mn^{3+} + \quad + ½O_2 \quad (5)$$

Symbol     represents an oxygen vacancy created by two $Mn^{3+}$ ions, as described in the equation. Our analysis reveals the presence of approximately 2% oxygen vacancies in the PMM [59]. In a previous study [60] on $Pr_2ZnMnO_6$, we found a 4% oxygen vacancy percentage. The presence of oxygen vacancies in the sample can be attributed to the fast-cooling rate during the calcination process.

### 2. Conductivity Analysis

The variation of AC conductivity ($\sigma_{ac}$) with frequency ($\omega$) is represented in a logarithmic scale graph [$\log \sigma_{ac}$ vs $\log \omega$] for different temperatures, as shown in Figure 9(a). The spectrum of $\sigma_{ac}$ values can be divided into two distinct regions. In the first region (Region I), where $\sigma_{ac}$ is independent of frequency, it exhibits the behavior of DC conductivity ($\sigma_{dc}$). In the second region (Region II), $\sigma_{ac}$ strongly depends on frequency. From the graph, it can be observed that the $\sigma_{ac}$ value increases with frequency in the high-frequency range for temperatures ranging from 303 K to 513 K. However, above 513 K, the $\sigma_{ac}$ value decreases with increasing frequency. To further investigate this behavior, the $\sigma_{ac}$ values are plotted against temperatures for two different frequencies [42 Hz (low frequency) and 5 MHz (high frequency)] in Figure 9(b). It is evident from the plot that the $\sigma_{ac}$ value increases with temperature for the low frequency of 42 Hz. However, for 5 MHz, the $\sigma_{ac}$ value initially increases with temperature, but above 513 K, it starts to decrease with increasing temperature. In Section 3.1.a, we have previously identified the phase transition in PMM with increasing temperature. The decrease in $\sigma_{ac}$ with temperature suggests metallic behavior in PMM at high energy values (high temperature and high frequency). This high energy value leads to the generation of a large number of free electrons in PMM. To confirm the metallic behavior in the high-temperature range, we employed the Lorentz-Drude Model [Equation 6] to fit the $\sigma_{ac}$ spectra.

$$\sigma_{ac} = \sigma_{dc}\omega_0^2 / [\omega^2 + \omega_0^2] \tag{6}$$

where $\omega_0 = \frac{1}{\tau}$, $\tau$ is the time constant. On the other hand, the low-temperature $\sigma_{ac}$ spectra fitted using power law [Equation 7].

$$\sigma_{ac} = \sigma_{dc} + C\omega^k \tag{7}$$

where $C$ and $k$ are the temperature-dependent constant and material-dependent parameter. The fitted parameters of $\sigma_{ac}$ are listed in Table 3. To determine the hopping model involved in the system, we plotted the $\sigma_{dc}$ values against the inverse of temperature ($10^3$/T) in the inset of Figure 9(b). The linear trend of the curve indicates the activation of the nearest neighbor hopping model at high-temperature regions [61, 62]. The activation energy value was calculated as 0.308 eV using Equation 8.

$$\sigma_{dc} = \sigma_0 e^{-(E_a/k_BT)} \tag{8}$$

where $\sigma_0$ is constant, $E_a$ is the activation energy, $k_B$ is the Boltzmann constant, and $T$ is the absolute temperature. The curve also exhibits a nonlinear behavior below 363 K, indicating the presence of variable-range hopping at low temperatures. To confirm that, we have plotted ln $\sigma_{dc}$ vs T$^{-0.25}$ graph in Figure 9 (c) using Equation 9.

$$\sigma_{dc} = Ae^{-[B/T]^{0.25}} \tag{9}$$

Here, A and B are constant parameters. The curve clearly shows a linear behavior at low temperatures. The inset of Figure 9(c) demonstrates that the activation energy ($E_a$) decreases at low temperatures, confirming the variable-range hopping model. Similar behavior has been reported in literature for materials such as $Pr_{0.8}Ca_{0.2}MnO_3$ [63]. Based on these results, we can conclude that $\sigma_{dc}$ undergoes a transition from nearest neighbor hopping to variable-range hopping around 363 K with a decrease in $E_a$.

3. **Electrochemical Activity**

The cyclic voltammetry (CV) curves of the PMM at various scan rates (5, 10, 20, 50, and 100 mV/s) are presented in Figure 10(a). The curves are plotted within a potential window of -0.2 V to 0.6 V (vs. Ag/AgCl) in a 1 M KOH aqueous electrolyte at room temperature. The CV curve exhibits two redox peaks: one peak associated with oxidation (~0.45 V vs Ag/AgCl) and the other with reduction (~0.25 V vs Ag/AgCl). This non-rectangular pseudocapacitive behavior of the PMM electrode indicates the occurrence of redox reactions between the $Mn^{3+}$ and $Mn^{4+}$ states at the electrode surface. The potential window in which the redox peaks are observed is based on previous studies on manganese-based electrodes [27]. Figure 10(c) shows that the peak current ($i_p$) increases with the scan rates ($v$) for both oxidation and reduction processes. This behavior suggests the quasi-reversibility of the electrode, indicating that the faradaic redox reaction controls the specific capacitance. To investigate the diffusion process involved in more detail, the logarithm of $i_p$ is plotted against the logarithm of $v$ for both oxidation and reduction processes. The slopes of the oxidation and reduction peaks are found to be 0.5725 and 0.5824, respectively, indicating that the diffusion-controlled intercalation process predominates over the pure surface capacitance process.

To illustrate the contribution of this process in PMM electrodes, we plot $i_p/v^{0.5}$ vs $v^{0.5}$ graph in the inset of Fig. 10 (c), with the help of Equation 10.

$$\frac{i_p}{v^{0.5}} = k_1 v^{0.5} + k_2 \qquad (10)$$

where $k_1$ and $k_2$ are parameters. The first parameter $k_1$ indicates that the $i_p$ contribution is due to surface capacitive process and the second parameter $k_2$ indicates the diffusion-controlled intercalation process. As can be seen from the figure, the $k_1$ and $k_2$ values are 0.014 (3.4%) and 0.401 (96.4) for oxidation peaks and those are 0.007 (3.5%) and 0.195 (96.5%) for reduction peaks, respectively. These results clearly demonstrate that the oxygen ion-controlled intercalation process is more significant than the pure surface capacitive process. The values of specific capacitance ($C_{sp}$) at various scan rate is calculated using Equation 11.

$$C_{sp} = \frac{\int i dV}{mv\Delta V} \qquad (11)$$

Here, $\int i dV$ is the area under the CV loop, m is the mass of the total active material in the electrode, v is the potential scan rate, and ΔV is the stability potential window. The PMM electrode exhibits a $C_{sp}$ value of 108 F/g at a scan rate of 2 mV/s. The $C_{sp}$ depends on the square root of the scan rate ($v^{0.5}$) as it represents a combination of the inner and outer surfaces of the electrode. Figure 10(c) depicts the graph of $C_{sp}$ vs $v^{0.5}$, while the inset shows the graph of log $C_{sp}$ vs log ν. The non-linear behavior observed in the $C_{sp}$ vs $v^{0.5}$ graph is attributed to an irreversible redox reaction [60]. As the scan rate approaches zero (ν → 0), the intercept of the log $C_{sp}$ vs log ν graph provides the maximum possible specific capacitance value of 257.57 F/g (i.e., $10^{intercept} = 10^{2.4109} = 257.57$).

The presence of approximately 4% oxygen vacancies in the PMM electrode, as mentioned in Section 3.1.c, plays a significant role in its pseudocapacitive behavior. The cyclic voltammetry (CV) curve clearly demonstrates this behavior. During the charging process, the oxygen vacancies within the PMM electrode are filled with oxygen ions from a 1 M KOH aqueous solution, facilitated by the Mn-O bond channels. This process is reversible, and the oxygen ions are released during the discharging process. Figure 10 (d) illustrates these charging and discharging processes, highlighting the intercalation of oxygen ions into the oxygen vacancies within the PMM electrode. This charging mechanism exhibits similarities to electrodes based on materials such as $LaMnO_{3\pm\delta}$ and $LaNiO_{3-\delta}$, as reported in references [27] and [58], respectively. The understanding of these processes and the presence of oxygen vacancies in the PMM electrode provide valuable information for the development of stable and high-performance perovskite-based oxygen-intercalated electrodes. These findings contribute to the advancement of oxygen-intercalated electrode technologies, enabling improved energy storage capabilities.

## 4. Conclusions

Crystal structure analysis and vibration studies indicate a phase transition of PMM. The thermal variation of the PMM indicates the phase transition from monoclinic to tetragonal around 500 K following the sequence $P2_1/n \rightarrow I4/m \rightarrow Fm\bar{3}m$. High-temperature Raman spectra of PMM show a significant feature of monoclinic – tetragonal phase transformation. The ac conductivity spectroscopy shows the semiconductor-to-metal transition in the PMM. The dc conductivity marks the transition from nearest neighbor hopping to variable-range hopping around 363 K with a decrease in $E_a$. The X-ray photoelectron spectroscopy of the Mn *2p* state shows the presence of oxygen vacancies in the PMM at room temperature. The PMM electrode exhibits intercalated pseudocapacitive nature with a maximum possible specific capacitance of 257.57 F/g. The charge storage process in the PMM electrode has been thoroughly reviewed and discussed, providing valuable insights for the development of stable and high-performance perovskite-based oxygen-intercalated electrodes.

**Conflict of interest**

There is no conflict of interest exists.

**Data availability statement**

The data that were used to support the findings of this study are available from the corresponding author upon reasonable request. Please contact the corresponding author directly to inquire about accessing the data.

Figures

***Figures:*** **Phase transition in oxygen-intercalated pseudocapacitor Pr$_2$MgMnO$_6$ electrode: A combined structural and conductivity analysis**

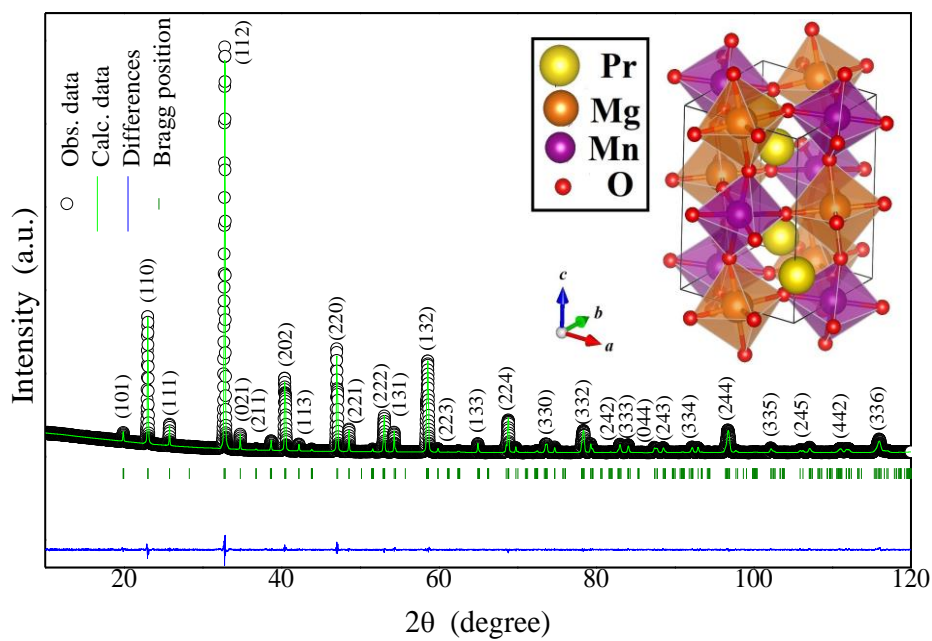

Figure 1

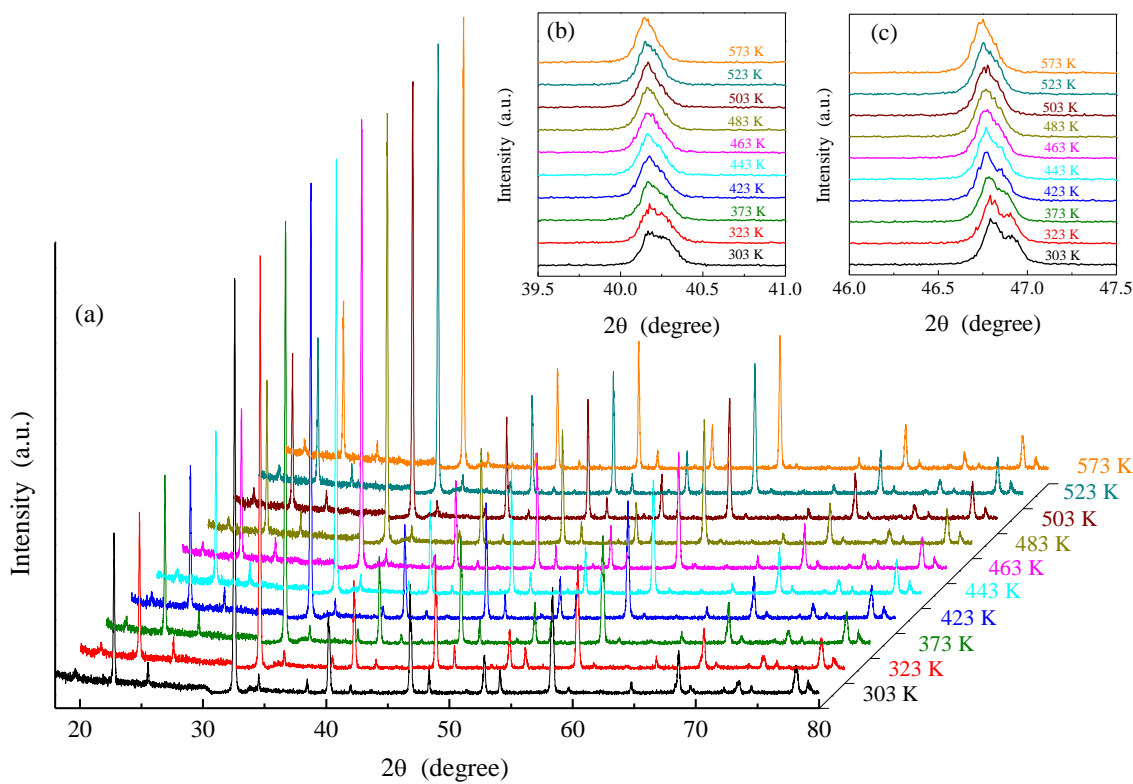

Figure 2

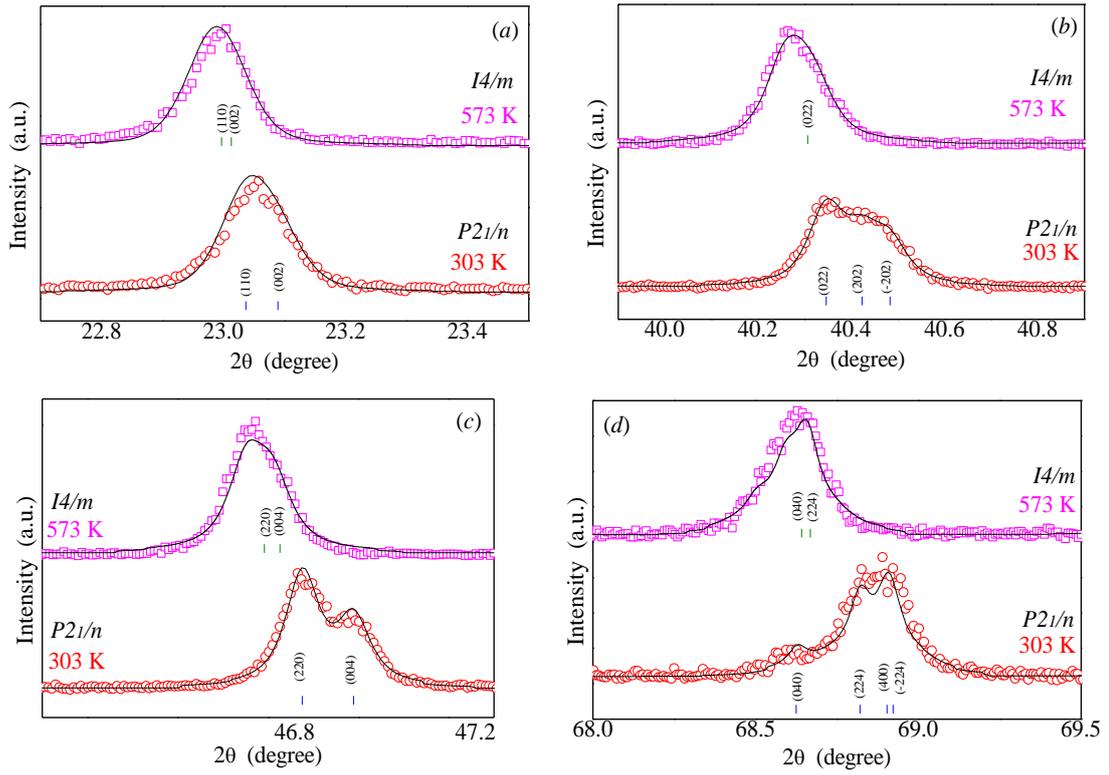

Figure 3

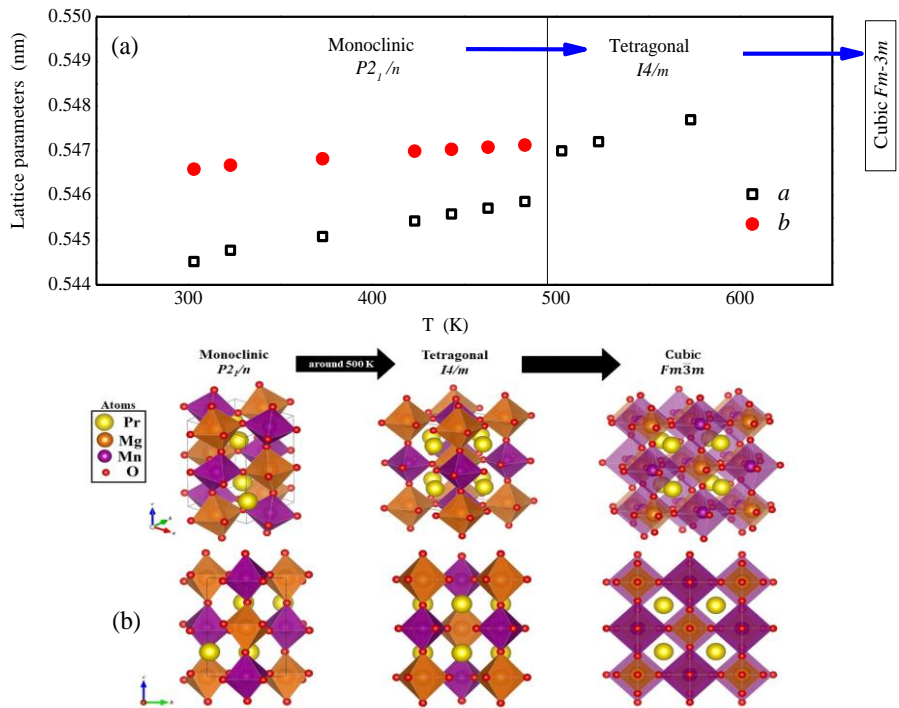

Figure 4

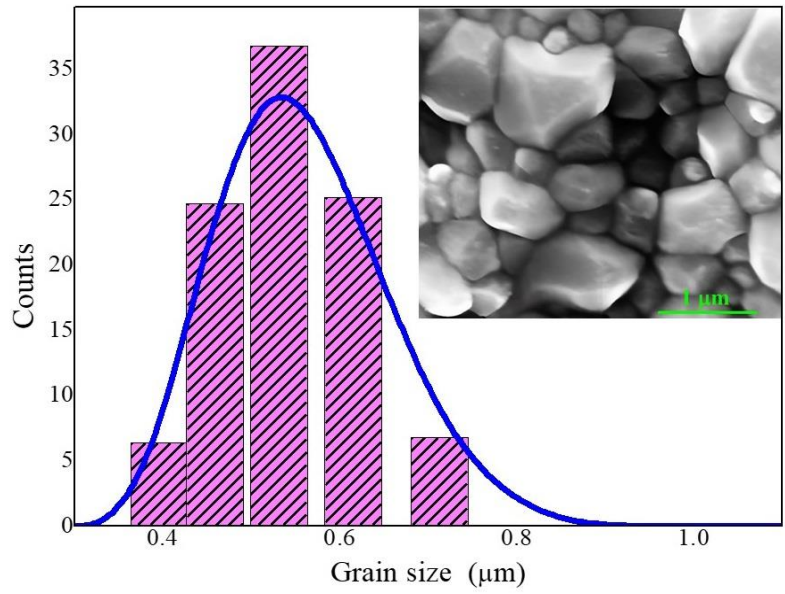

Figure 5

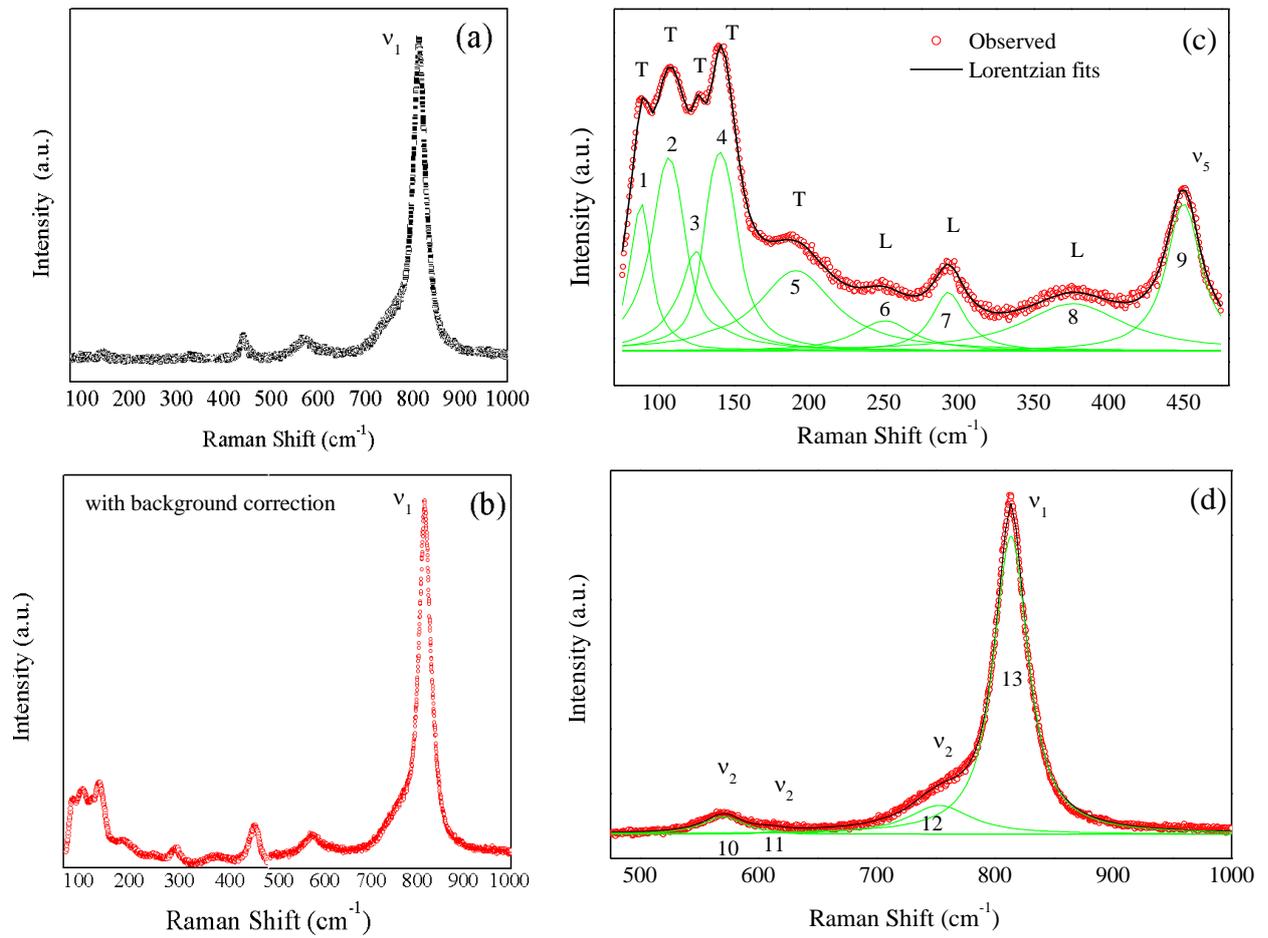

Figure 6

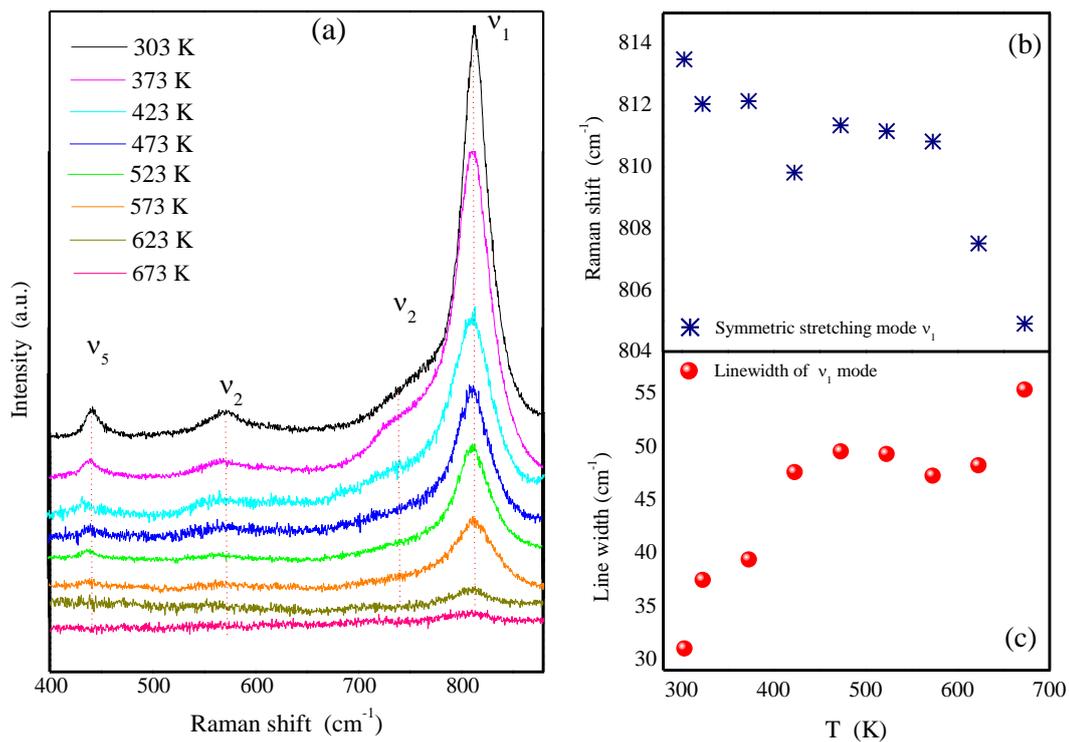

Figure 7

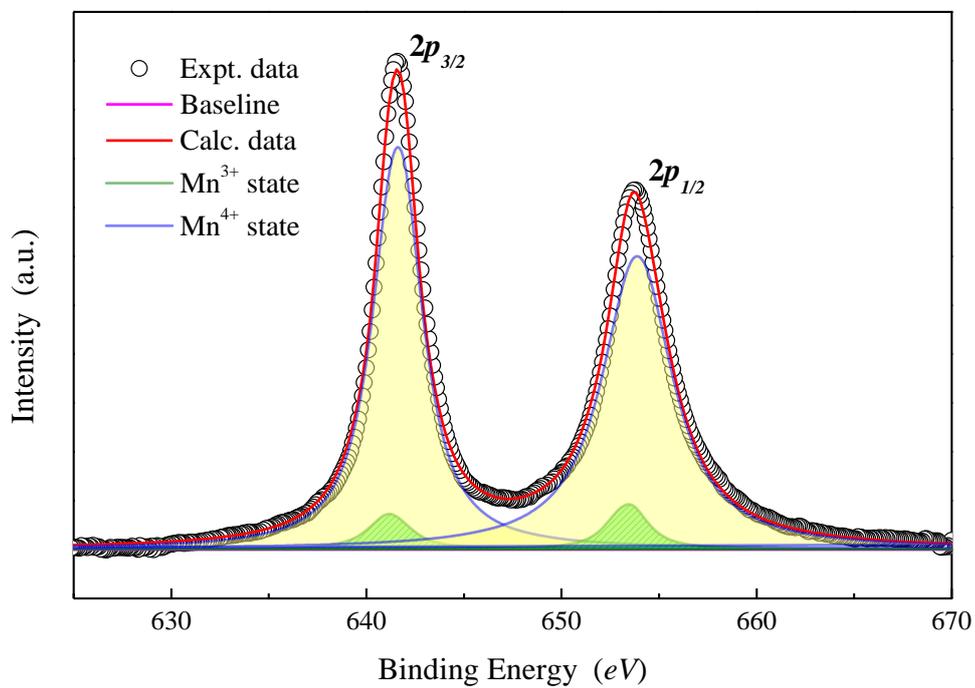

Figure 8

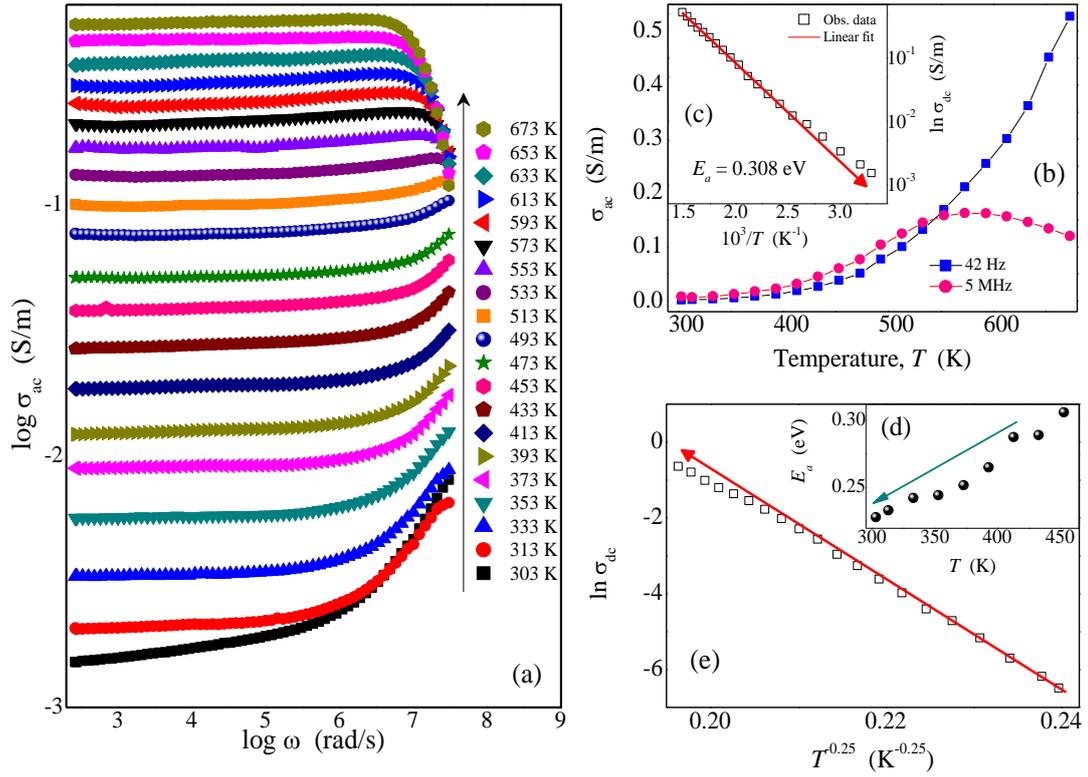

Figure 9

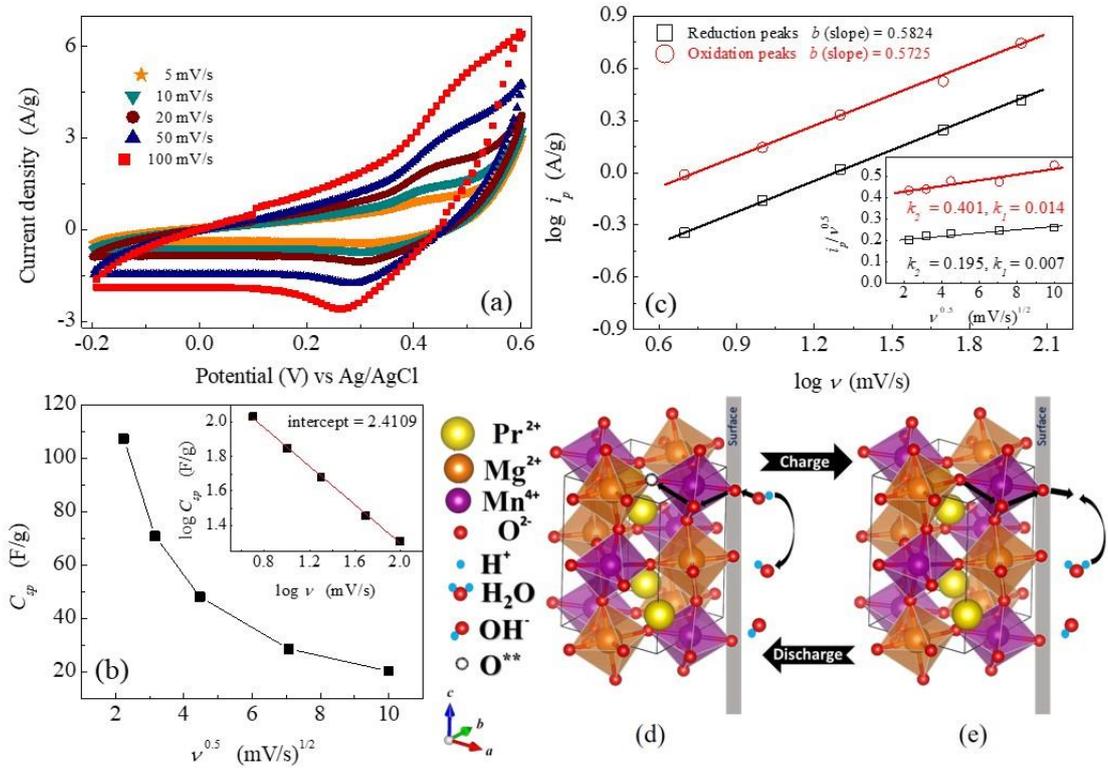

Figure 10

*Figure captions:*

Figure 1: Rietveld analysis results of XRD data for PMM at RT. The appendix shows the unit cell of PMM as obtained from Rietveld analysis.

Figure 2: The XRD patterns are collected from RT to 573 K for PMM.

Figure 3: The Rietveld refinement of the XRD pattern of PMM at RT and 573 K for different $2\theta$ values. The values of $2\theta$ from (a) $22.6^0$ - $26^0$, (b) $40^0 – 41^0$, (c) $46.5^0 – 47.5^0$, and (d) $68^0 – 70.4^0$.

Figure 4: Variation of unit cell parameters (a) and phase (b) of PMM with temperature.

Figure 5: Grain size distribution curve of PMM calcined at 1573 K. Appendix shows SEM image of PMM pellets.

Figure 6: (a) The RT unpolarized Raman spectrum of PMM. (b) The RT Raman spectrum of PMM with background correction. The entire deconvoluted Raman spectrum is separated for a better view into two ranges: (c) $50 – 480$ cm$^{-1}$, and (d) $480 – 1000$ cm$^{-1}$ ($\times 10^2$ m$^{-1}$). (Experimental data are red open circles, while green solid lines represent phonon modes adjusted by Lorentzian curves).

Figure 7: (a) Temperature-dependent Raman spectra for PMM. The thermal variation of (b) the Raman shift and (c) the linewidth of the symmetric stretching mode ($\upsilon_1$).

Figure 8: The XPS spectrum of Mn 2*p* state for PMM.

Figure 9: (a) Frequency dependence of the ac conductivity ($\sigma_{ac}$) at different temperatures for PMM. (b) Temperature dependence of $\sigma_{ac}$ at 42 Hz and 5 MHz. The appendix shows the dc conductivity ($\sigma_{dc}$) plotted using the nearest-neighbor hopping model. (c) The $\sigma_{dc}$ using variable range hopping model. The appendix shows the variation of activation energy (E*a*) with the temperature.

Figure 10: (a) The CV curves of PMM electrode at several scan rates, (b) The variation of C$_{sp}$ with $\nu^{0.5}$ and the inset shows log C$_{sp}$ vs log $\nu$ plot, (c) log i$_p$ vs log $\nu$ plot and the inset shows i$_p$/$\nu^{0.5}$ vs $\nu^{0.5}$ plot. Schematic diagram of (d) oxygen intercalation into PMM and (e) oxygen deintercalation from PMM.

**Tables: Phase transition in oxygen-intercalated pseudocapacitor Pr$_2$MgMnO$_6$ electrode: A combined structural and conductivity analysis**

Table – 1

Refined structure parameters of PMM at 303 K and 573 K.

| Temperature 303 K<br>Monoclinic P2$_1$/n | | | | | Temperature 573 K<br>Tetragonal I4/m | | | | |
|---|---|---|---|---|---|---|---|---|---|
| Cell parameters:<br>$a$ =5.44826 (5) Å, $b$ = 5.46948(4) Å,<br>$c$ = 7.70182(5) Å, $\beta$ = 89.9063° | | | | | Cell parameters:<br>$a$ =5.4647 (5) Å,<br>$c$ = 7.7230(6) Å. | | | | |
| Atom | Wyckoff site | $x$ | $y$ | $z$ | Atom | Wyckoff site | $x$ | $y$ | $z$ |
| Pr | 4e | -0.0051 | 0.0340 | 0.2512 | Pr | 4d | 0 | 0.5 | 0.25 |
| Mg | 2c | 0.5 | 0 | 0 | Mg | 2a | 0 | 0 | 0 |
| Mn | 2d | 0.5 | 0 | 0.5 | Mn | 2b | 0.5 | 0.5 | 0 |
| O1 | 4e | 0.0680 | 0.4868 | 0.2468 | O1 | 4e | 0 | 0 | 0.2858 |
| O2 | 4e | 0.7118 | 0.2873 | 0.0336 | O2 | 8h | 0.2026 | 0.3302 | 0 |
| O3 | 4e | 0.2093 | 0.2052 | 0.9691 | | | | | |
| R$_p$ = 5.21, R$_{wp}$ = 4.95, R$_{exp}$ = 2.71, $\chi^2$= 3.338. | | | | | R$_p$ = 2.91, R$_{wp}$ = 3.89, R$_{exp}$ = 2.52, $\chi^2$= 2.39. | | | | |

Table – 2

Correlation between the Raman modes of Monoclinic, Tetragonal, and Cubic Phases of PMM.

| | $P2_1/n$ | $I4/m$ | $Fm\bar{3}m$ |
|---|---|---|---|
| | $12A_g$ ⟶ $3A_g$ ⟶ | $A_{1g}$ | |
| | ↗ $3B_g$ ⟶ | $E_g$ | |
| | $12B_g$ ⟶ $6E_g$ ⟶ | $2T_{2g}$ | |
| $\Gamma^g =$ | $12A_g + 12B_g$ | $3A_g + 3B_g + 6E_g$ | $A_{1g} + E_g + 2T_{2g}$ |

Table – 3

Fitted parameters of $\sigma_{ac}$ for different temperatures.

| Fitted using Lorentz-Drude Model | | |
|---|---|---|
| Temperature (K) | $\sigma_{dc}$ (S/m) | $\omega_0 = \frac{1}{\tau}$ (s$^{-1}$) ($\times 10^6$ Hz) |
| 673 | 0.568 | 18.5 |
| 653 | 0.485 | 23.5 |
| 633 | 0.405 | 28 |
| 613 | 0.348 | 31 |
| 593 | 0.295 | 39 |
| 573 | 0.248 | 49 |
| 553 | 0.199 | 70 |
| 533 | 0.167 | 80 |
| 513 | 0.13 | 90 |

| Fitted using Power Law | | | |
|---|---|---|---|
| Temperature (K) | $\sigma_{dc}$ (S/m) | Temperature-dependent constant, $C$ ($\times 10^{-9}$) | Material-dependent constant, $k$ |
| 493 | 0.08 | 1.4 | 0.97 |
| 473 | 0.054 | 1.5 | 0.96 |
| 453 | 0.04 | 2.2 | 0.93 |
| 433 | 0.0288 | 2.45 | 0.91 |
| 413 | 0.0194 | 5 | 0.85 |
| 393 | 0.0128 | 5 | 0.84 |
| 373 | 0.00912 | 9.9 | 0.79 |
| 353 | 0.00575 | 14.2 | 0.76 |
| 333 | 0.00337 | 20 | 0.73 |
| 313 | 0.00211 | 26 | 0.71 |
| 303 | 0.00182 | 37 | 0.70 |